\documentstyle[aps,epsfig,floats]{revtex}
\draft  
\begin{document}
\twocolumn[\hsize\textwidth\columnwidth\hsize\csname
@twocolumnfalse\endcsname
\title{Signatures of the efficiency of solar nuclear reactions
in the neutrino experiments}

\author{H. Schlattl}
\address{Max-Planck-Institut f\"ur Astrophysik, Karl-Schwarzschild-Str. 1, 
85740 Garching, Germany}
\author{A. Bonanno$^{1,2}$, L. Patern\`o$^1$}
\address{$^1$Istituto di Astronomia dell'Universit\`a, 
Viale A.Doria 6, 95125 Catania, Italy} 
\address{$^2$ INFN, sezione di Catania,
Corso Italia 57, 95128 Catania, Italy}
\date{\today}

\maketitle

\begin{abstract}
In the framework of the neutrino oscillation scenario, we discuss the
influence of the uncertainty on the efficiency of the neutrino
emitting reactions $^{1}{\rm H}(p,{e}^+\nu_{e})^{2}{\rm H}$ and
$^7{\rm Be}({\rm p},\gamma)^8{\rm B}$ for the neutrino oscillation
parameters. We consider solar models with zero-energy astrophysical 
S-factors $S_{11}$ and $S_{17}$ varied within nuclear physics uncertainties, 
and we test them by means of helioseismic data.
We then analyse the neutrino mixing parameters and recoil electron spectra 
for the presently operating neutrino experiments and we predict the results which 
can be obtained from the recoil electron spectra in 
SNO and Borexino experiments. We suggest that it should be
possible to determine tight bounds on $S_{\rm 17}$ from the
results of the future neutrino experiment, in the case of
matter-enhanced oscillations of active neutrinos.
\end{abstract}
\pacs{PACS numbers: 26.65.+t, 14.60.Pq, 96.60.Ly, 25.60.Pj}
]

\section{Introduction}
The solar neutrino experiments - Homestake (HM), Kamio\-kande (K), 
GALLEX, SAGE,
and Super-Kamiokande (SK) -  have shown the existence of robust quantitative
differences between the experiments and the combined predictions of
minimal standard electroweak theory and stellar evolution theory. On
the other hand, the latter is nowadays in significant agreement with 
the constraints posed by helioseismology
and thus we can consider the possibility that the
neutrinos have properties other than those included in the standard electroweak model.
The Mikheyev-Smirnov-Wolfenstein~\cite{msw} (MSW) matter-enhanced
oscillation and vacuum (``just-so'') oscillation~\cite{vo} (VO)
provide an explanation of the neutrino deficit, although it is not yet clear
which mechanism produces the required suppression.  Due to the
increasing accuracy of the results of the present and 
future neutrino experiments, 
it is interesting to investigate the effects of the uncertainty in solar
physics parameters on the solar neutrino oscillation scenarios.

We study how the allowed regions in the parameter
space of the two-flavour oscillations are modified when $S_{11}$ and
$S_{17}$, the astrophysical zero energy S-factors of the reactions
$^1{\rm H}(p,e^+\nu_{\rm e})^2{\rm H}$ and $^7{\rm Be}(p,\gamma)^8{\rm
B}$, are changed within the ranges derived from the 
nuclear physics calculations and experiments,
using up-to-date solar models. 

The efficiency of the first reaction determines, through $S_{11}$, the
evolution of the chemical composition in the Sun and its hydrostatic
structure. Since the meteoritic age of the Sun is fairly well
known~\cite{BP95}, any modification of $S_{11}$ changes the
present central abundance of hydrogen and hence the behaviour of
the adiabatic sound speed which can also be determined 
by helioseismic p-mode data inversion.  
Most of the astrophysical S-factors of the relevant
nuclear reactions in the Sun are determined from measurements in the
laboratory at higher energies, extrapolated down to zero
energy. However, due to the very rare event rate of $^1{\rm
H}(p,e^+\nu_{\rm e})^2{\rm H}$ at high energies (1 reaction in $\sim
10^6$ years at 1~MeV for a proton beam of 1 mA~\cite{PR91}) this
procedure is not applicable to $S_{11}$ and its estimation must be
obtained from standard weak-interaction theory~\cite{Kam94}.  The
latest suggested value~\cite{Adel98} is $S_{11}=4.00\;10^{22}~
{\rm keV\,barn}$ with an uncertainty of $\simeq\pm\,2.5\,\%$ at
$1\,\sigma$.  This is of the same order as the uncertainty of the free
neutron decay time, which is linked to the ratio of the axial-vector
to the Fermi weak-coupling constants.

The reaction $^7{\rm Be}(p,\gamma)^8{\rm B}$ produces the dominant signal in the HM, 
SK and Sudbury Neutrino Observatory (SNO) neutrino
experiments. Unfortunately $S_{\rm 17}$ is one of the most poorly
known quantity of the entire nucleosynthesis chain which leads to the
${\rm^8B}$ formation. The reaction cross-section is measured down to
$134~{\rm keV}$ with large statistical and systematic errors which
dominate the uncertainty in the determination of the astrophysical
factor at low energies~\cite{Tur93}. 
\cite{Adel98} recently quoted
$S_{17}=19^{+4}_{-2}~{\rm eV\,barn}$ at
$1\sigma$, suggesting a conservative value of $S_{17}$ in the range
$15~{\rm eV\,barn}$ to $27~{\rm eV\, barn}$ with an error of
$\simeq\pm 30\%$ at $3\sigma$.  Any change in $S_{\rm 17}$ affects
only the $^8{\rm B}$ neutrino flux, $\phi_{\nu}(^8{\rm B})$, and leaves
all the other relevant quantities of the solar model, such
as the sound speed profile and the neutrino fluxes produced in the other
reactions, unaltered~\cite{Pat97}.

The value of $S_{11}$ influences indirectly the total $\phi_{\nu}(^8{\rm B})$
which is quite sensitive to the central temperature of the Sun $T_{\rm c}$
($\phi_\nu(^8 {\rm B})\propto S_{17} {T}_{\rm c}^{24}$~\cite{newb}.).
In fact, a change in $S_{11}$ determines a change in both the total
$pp$-neutrino flux, $\phi_{\nu}(pp)$, and $T_{\rm c}$, 
being $\Delta T_{\rm c}/T_{\rm c}\simeq -0.15\, \Delta S_{11}/S_{11}$ and
$\phi_\nu(pp)\propto T^{-1}_{\rm c}$. We therefore constrain $S_{11}$ with
the help of helioseismology, in order to reduce its influence on the
total $\phi_{\nu}(^8{\rm B})$ uncertainty.  Nonetheless, 
the greatest uncertainty
in this flux still remains the measurement of $S_{17}$.

Since the efficiency of $^1{\rm H}(p,e^+\nu_{\rm e})^2{\rm H}$ 
influences mainly the structure of the solar model and the
neutrino rates, whereas the situation is opposite for the strength of
$^7{\rm Be}(p,\gamma)^8{\rm B}$, we have considered the following
cases: 1) $S_{17}$ standard and $S_{11}$ varied; 2) $S_{11}$ 
standard and $S_{17}$ varied.  Here by {\it standard} we denote the
most favoured values for $S_{17}$ and $S_{11}$ suggested
by~\cite{Adel98} and by {\it varied} we mean a conservative range of
variations allowed at $\sim 99\%$ confidence level. All the other reaction rates,
such as $S_{33}$ and $S_{34}$, are left unaltered to their 
standard values as given in~\cite{Adel98}.
In Section II we
investigate case 1) by using helioseismic data in order to obtain more
stringent limits on the ``unsuppressed'' total $\phi_{\nu}(pp)$.

\begin{table}[b]
\caption{\label{rates}
Solar neutrino event rates with $1\sigma$ errors. In the theoretical
errors the $S_{\rm 17}$-uncertainty is removed.}
\smallskip
\begin{tabular}{llc}
Experiment & Data $\pm$ (stat)$\pm$(syst.)\tablenotemark[1] & theor.~err.\tablenotemark[2]\\
\noalign{\vskip3pt\hrule\vskip3pt}
HM & $2.56\pm 0.16 \pm 0.15~{\rm SNU}$ & 13.1\% \\
SAGE & $69.9^{+8.0\,+3.9}_{-7.7\, -4.1}~{\rm SNU}$ & 5.8\% \\
GALLEX & $76.4\pm6.3^{+4.5}_{-4.9}~{\rm SNU}$ & 5.8\% \\
SK & $2.44\pm0.05^{+0.09}_{-0.07}~{\rm 10^6cm^{-2}s^{-1}}$& 14\% \\ 
\noalign{\vskip3pt\hrule\vskip3pt}
GALLEX+SAGE & $72.4\pm6.6~{\rm SNU}$& 5.8\%\\
\end{tabular}
\footnotesize
\tablenotemark[1]{H.~Minakata and H.~Nunokawa, hep-ph/9810387 (1998)}\\
\tablenotemark[2]{Derived from~\cite{Bah98}}
\normalsize
\end{table}

The behaviour of the neutrino mixing parameters $\Delta m^2$ and
$\sin^22\theta$ as a function of $S_{17}$ is presented in Section
III. The mixing parameters are obtained through $\chi^2$-fits by using the recent
results of HM, GALLEX, SAGE and SK experiments as shown in
Table~\ref{rates}.  We consider MSW and VO transitions into active
(non-sterile) and sterile neutrinos as well. Previous analyses in this
direction have been carried out by other authors which used different
approaches and considered an arbitrary $\phi_{\nu}(^8{\rm B})$
~\cite{lan} as an additional free parameter.  
In our calculations the Earth regeneration
effect is included and the exact evolution equation for the neutrino
mixing is solved numerically without resorting to analytical
approximations.

In Section~IV, the first and second moments of the recoil electron spectra in
SK are calculated for the best-fit values of ${\sin}^{2}2\theta$ and
$\Delta {\rm m}^{2}$ obtained in the previous section, and we
discuss the possibility of considering $S_{17}$ as a free parameter in
the analysis of the forthcoming data from both SNO and Borexino
experiments. It is shown that a determination of the lower and upper
limits on the $S_{17}$ values 
can be derived from the measurement of the charged current to neutral current
relative ratio (CC/NC) in SNO. Section V is devoted to the conclusions.

\section{solar models}  
The solar models were computed by using the latest version of the
GArching SOlar MOdel (GARSOM) code, which originates from the
Kippenhahn stellar evolution program~\cite{KWH67}. Its numerical and
physical features are described in more details in~\cite{Sch97}. In
particular, it uses the latest OPAL-opacities~\cite{OPOPAC} and
equation of state~\cite{OPEOS} and it takes into account microscopic
diffusion of hydrogen, helium and heavier elements (e.g.~C, N, O). The
diffusion constants are calculated by solving Burgers' equation for a
multicomponent fluid via the routine described in ~\cite{Thoul}. The standard
values of the reaction rates are taken from~\cite{Adel98}. In the present
version the equations for nuclear network and diffusion are solved
simultaneously.  We follow the evolution of the models from ZAMS to
an age of 4.6 Gyr.  The metal abundances are taken from~\cite{Gre93}.
The convection is described by the mixing length theory~\cite{MLT}.
Unlike previous work~\cite{Sch97}
where models of solar atmosphere were used, here we consider an 
Eddington atmosphere for the outer boundary conditions since we focus
our attention on processes occurring in the deep interior, where
the exact stratification of the atmosphere has almost no influence. 

\begin{figure}[ht]
\hbox to\hsize{\hss\epsfxsize=8cm\epsfbox{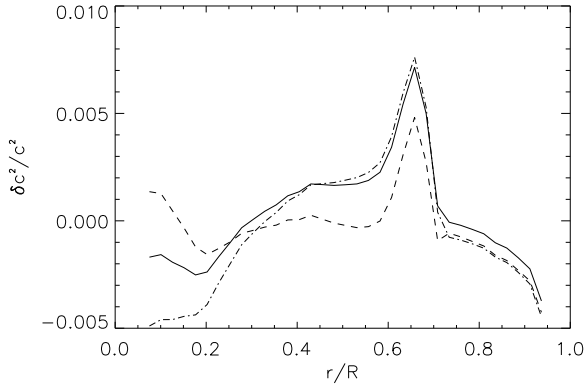}\hss}
\caption{Difference in sound-speed profiles of various solar
models. The solid line is our standard solar model, the dash-dotted
line is obtained for $S_{11}=3.89\;10^{-22}~{\rm keV\,barn}$, and the
dashed one for $S_{11}=4.20\;10^{-22}~\rm{keV\,barn}$.
\label{cspeed}}
\end{figure}

A comparison of the present model with other up-to-date standard solar 
models is given in~\cite{Sch99}. 
In Fig.~\ref{cspeed} we show the behavior of sound speed in our standard solar
model as compared with the seismic model derived by S.~Basu and
J.~Christensen-Dalsgaard by inverting the GOLF+MDI data ~\cite{silv1}.  
The values of some basic quantities of our models are
summarized in Table~\ref{models}. These values refer to the computation
of different solar models with $S_{11}$ varied
within the extreme cases of $3.89\;10^{-22}~{\rm keV\,barn}$
and $4.20\;10^{-22}~{\rm keV\,barn}$, and $S_{17}$ kept at the
standard value (case 1).  
Each model with a given value of $S_{11}$ has been obtained by
following the whole evolution and adjusting 
the mixing length, initial helium abundance, 
and chemical composition to fit solar
luminosity, effective temperature and the surface value of 
$Z/X=0.0245$~\cite{Gre93}. All other 
input physics, like opacity or the equation of state, is the same
for all the models.
In particular, the  luminosity and effective temperatures of the models differ from solar 
luminosity and effective
temperature less than $10^{-4}$ for all the models considered. 

\begin{table}[ht]
\caption{\label{models} Solar models with different values of
$s_{11}$= $S_{11}/(10^{-22}~{\rm keV\,barn})$, the \emph{standard}
model has $s_{11}$=4.00.}
\smallskip
\begin{tabular}[2]{ccccccc}
$s_{11}$ & ${\rm T_{c}}$ & ${\mu_{c}}$ & $R_{cz}/R_\odot$ &
$\phi(^8{\rm B})$ & GALLEX & HM\\ 
& $(10^7{\rm K})$ & &  & $({\rm
cm}^{-2} {\rm s}^{-1})$ & (SNU) & (SNU)\\
\noalign{\vskip3pt\hrule\vskip3pt} 3.89 & 1.578 & 0.860 & 
0.715 & 5.54~$10^6$ & 131.8 & 8.2 \\ 4.00 & 1.574 & 0.859 & 
0.713 & 5.16~$10^6$ & 129.5 & 7.7 \\ 4.10 & 1.567 & 0.858 & 
0.712 & 4.85~$10^6$ & 127.5 & 7.3 \\ 4.20 & 1.563 & 0.857 & 
0.711 & 4.56~$10^6$ & 125.6 & 6.9 \\
\end{tabular}
\end{table}

The production region of the $pp$ neutrinos extends up to
$r<0.3\,R_\odot$. This region is within the reach of the low order
$p$-modes. It can be of interest to verify to which extent
the uncertainty in the theoretical calculations of $S_{11}$ can be constrained by
helioseismic data. In order to investigate this
possibility we have compared the sound speed profile of solar models
with different $S_{11}$ with the sound speed profile derived 
from helioseismic data inversion in ~\cite{silv1}.
The result is shown in ~Fig.~\ref{cspeed} where it appears that a model 
with the highest $S_{11}$ better reproduces the internal stratification.
 
A different method is the forward approach where small differences in
frequencies of low order modes are compared. The small spacing differences
$\delta \nu_{n,l} =\nu_{n,l} -\nu_{n-1,l+2}$, for $l=0$ and $l=1$, are in fact
highly sensitive to the sound speed gradient in the very central region
of the Sun. For this purpose we have then used a weighted average of the first 
144 days of MDI and of 8 months GOLF data~\cite{golf}
for $l=0,1,2,3$ and $n$ from $10$ up to $26$.
We have thus calculated $\delta\nu_{n ,l}$ for $l=0$ and $l=1$ relative to
solar models with different $S_{11}$.
\begin{figure}[ht]
\hbox to\hsize{\hss\epsfxsize=8cm\epsfbox{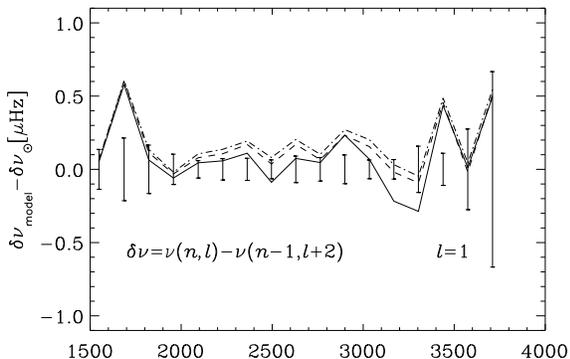}\hss}
\caption{Differences in small spacing frequency differences for
various solar models. The solid line is obtained for
$S_{11}=4.20\;10^{-22}~{\rm keV\,barn}$, the dash-dotted line for
$S_{11}=3.89\;10^{-22}~{\rm keV\,barn}$, and the dashed line for the
standard case $S_{11}=4.00\;10^{-22}~{\rm keV\,barn}$.
\label{sms}}
\end{figure}

From an ispection of Fig.~\ref{sms} it appears that, for $l=1$, the
model with the highest $S_{11}$ seems to approach more closely
the real Sun (a similar conclusion is obtained for $l=0$). 
This is consistent with the results of
secondary inversions for the temperature profile where it has been
estimated that $S_{11}=(4.15\pm 0.25)\;10^{-22}~{\rm
keV \, barn}$ ~\cite{Antia98}.  
Since both the inverse and forward helioseismic approach indicate 
that higher values of $S_{11}$ seem more favoured, we are allowed 
to conclude that the total $\phi_{\nu}(pp)
\propto S_{11}^{0.14}S_{33}^{0.03}S_{34}^{-0.06}$ can be
considered as bounded from below at the value
\[5.93\;10^{10} {\rm cm^{-2} s^{-1}}\leq \phi(pp)\]
from helioseismic data.

The greatest uncertainty in the neutrino flux predicted by solar
models comes from the poorly known $^8{\rm B}$-neutrinos, whose flux
is mainly determined by the reaction
rate of $^7{\rm Be}(p,\gamma)^8{\rm B}$, the first reaction of the $pp$III-subcycle.
This subcycle contributes by only 0.01\% to the total energy production of the 
$pp$-cycle though it is responsible for the emission
of the most energetic neutrinos produced in this subcycle. 
Its contribution has practically no influence on the solar structure, 
thus excluding any possibility of producing signatures 
in the helioseismic frequencies. For the computations that follow we keep
$S_{11}$ fixed at its standard value,
and we vary $S_{17}$ within the allowed ``conservative'' range.

\section{results for neutrino oscillation parameters}
In this section we present the results obtained from the total
rates in the GALLEX/SAGE, HM and SK detectors (Table~\ref{rates}) for
our modified solar model introduced in the previous section.  We have
calculated the allowed parameter space ($\Delta m^2, {\sin ^2
2\theta}$) for neutrino oscillations in the two-flavour case, taking the
theoretical errors from~\cite{Bah98}. As we study the
influence of $S_{17}$ on the oscillation parameters we remove its
contribution from the total theoretical uncertainty.
\begin{figure}[ht]
\hbox to\hsize{\hss\epsfxsize=9cm\epsfbox{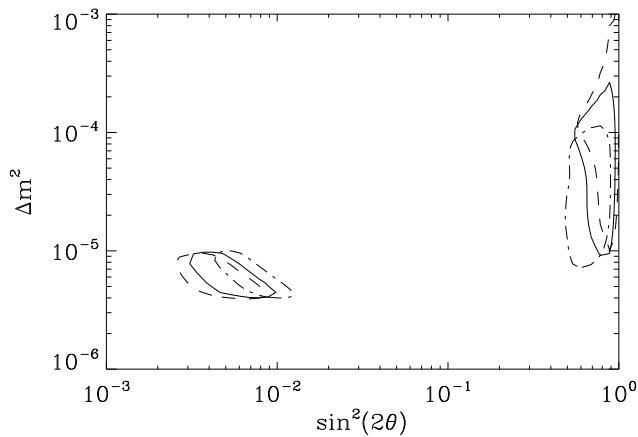}\hss}
\caption{Allowed regions (95\% C.L.) of neutrino mixing parameters in
a two flavour case for solar models with different cross sections of
$^7{\rm Be}(p,\gamma)^8{\rm B}$ ($S_{\rm 17} = 17$ dashed, 19 solid
and 23 dash-dotted, in units of ${\rm eV\,barn}$).
\label{plane}}
\end{figure}
For the calculation of the MSW-effect we piecewise linearize the
density profile of the respective solar models, and  the evolution
equations for neutrino oscillations are then integrated by using the
exact solution on each linear part. We also include the average
earth-regeneration effect~\cite{Earth}. 
 Since the models with different values of $S_{17}$ predict a different
$^8$B-neutrino flux, the expected event rate changes for SK and 
also for GALLEX/SAGE and HM. Thus, different
conversion probabilities are needed for each value of $S_{17}$ in order 
to explain the
measured rates in these experiments. This leads to different
confidence regions in the $\sin^22\theta-\Delta m^2$-plane (see
Fig.~\ref{plane}). 
%
%

\begin{figure}[ht]
\hbox to\hsize{\hss\epsfxsize=9cm\epsfbox{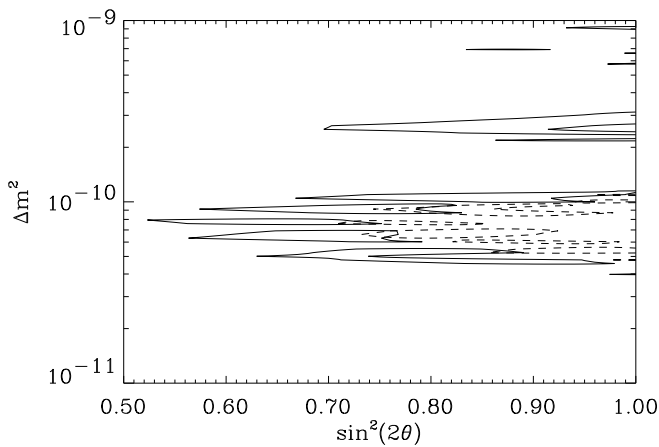}\hss}
\caption{Allowed region (95\%) for VO as $S_{\rm 17}$ is varied
($S_{17}=16$ solid, $S_{17}=23$ dashed)
\label{sno2}}
\end{figure}

The general trend in the small mixing angle (SMA) solution shows 
that an increase of $S_{17}$
shifts the mixing towards larger angles, while keeping the mass
difference almost constant. Similar trend can also be noted for the VO
case (Fig.~\ref{sno2}).  In the large mixing angle (LMA) 
solution both the mass difference and
the mixing angle decrease with increasing $S_{17}$.

\begin{figure}[t]
\hbox to\hsize{\hss\epsfxsize=9cm\epsfbox{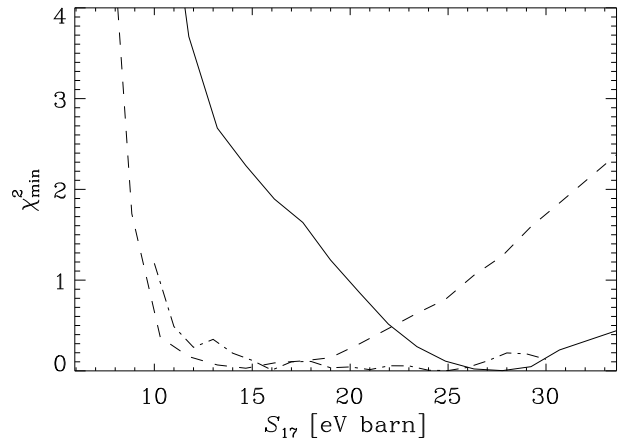}\hss}
\caption{The minimal $\chi^2$-values with varying strength of $^7{\rm
Be}(p,\gamma)^8{\rm B}$ for the LMA (solid), SMA (dashed) and
VO-solution (dash-dotted line).
\label{sno_chi}}
\end{figure}

The results shown in Fig.~\ref{plane} and Fig.~\ref{sno2}
indicate that if on one hand there are always three possible well 
separated solutions of VO, SMA and LMA, on the other hand
it is difficult to disentangle additional effects in each of the solutions for the
present experimental status, since $\chi^2$ has rather shallow
minima (Fig.~\ref{sno_chi}).
\begin{table}[hb]
\caption{\label{matter} Best-fit solutions for the total event rates
in Table I. The first two column refers to SMA solution, the second to
LMA and the last ones to VO. $S_{17}$ is given in ${\rm eV\,barn}$}
\smallskip
\begin{tabular}[2]{cccccccc}
$S_{17}$ & $\Delta m^2$ & $\sin^2(2\theta)$ & $\Delta m^2$ &
$\sin^2(2\theta)$ & $\Delta m^2$ & $\sin^2(2\theta)$ \\
\noalign{\vskip3pt\hrule\vskip3pt} 
15 & $5.2\;10^{-6}$ & $4.2\;10^{-3}$ & $2.7\;10^{-4}$ & 0.88 & 
$1.1\;10^{-10}$ & 0.88\\ 
17 &$5.2\;10^{-6}$ & $6.1\;10^{-3}$ & $8.5\;10^{-5}$ & 0.88 &
$1.1\;10^{-10}$ & 0.93\\ 
19 & $5.3\;10^{-6}$ & $6.5\;10^{-3}$& $7.4\;10^{-5}$ & 0.82 & 
$9.1\;10^{-11}$ & 0.78\\ 
23 & $5.2\;10^{-6}$ & $8.8\;10^{-3}$ & $2.1\;10^{-5}$ & 0.69 & 
$6.6\;10^{-11}$ & 0.85\\ 
27 & $5.3\;10^{-6}$ & $1.0\;10^{-2}$ & $1.6\;10^{-5}$ & 0.57 &
$8.7\;10^{-11}$ & 0.95\\
\end{tabular}
\end{table}
A constraint on $S_{17}$ at $1\sigma$  ($\chi^2$$-$$\chi^2_{\rm min}$=1 in 
Fig.~\ref{sno_chi}) 
can be obtained from the $\chi^2$ analysis of the total neutrino rate in the
case of SMA solution which gives $9\leq S_{17}\leq 25$
and it leads to the following constraint on the $\phi_{\nu}(^8{\rm B})$
\[0.6 \leq f_{\nu_n}(^8{\rm B}) \leq 1.8 ~~~~~ (2\sigma), \]
where $f_{\nu_n}(^8{\rm B})$ is the normalized neutrino flux
$\phi_{\nu}(^8{\rm B})/\phi_{\nu}(^8{\rm B})|_{standard}$.

We have also analysed the case of a non-standard
$S_{11}$ concluding that, if the range of variation is limited by both
helioseismology and nuclear physics uncertainties, the differences in
the best-fit solutions are not very significant.
Unfortunately, at the present time it is not clear which kind of
oscillation mechanism is responsible for the neutrino suppression.
Additional information should be available from the future data of SK,
Borexino and SNO experiments.

\section{Future data and experiments}
%
%
In the following sections the expected forthcoming data for SK,
Borexino and SNO are summarized. We focus (i) on the ability of these
experiments to identify the oscillation mechanism (LMA, SMA or VO))
and (ii) on what is expected to be measured in these detectors using
solar models with different values of $S_{17}$ and taking into account
the present data of GALLEX/SAGE, HM and SK.
%
%
\subsection{Super-Kamiokande}

Recently, the SK-collaboration published first data about the zenith
angle dependence~\cite{SKzenit} of neutrino flux and electron recoil energy
spectrum~\cite{SKele} which seem to disfavour any of the above
investigated solutions. However the present statistics and detector 
threshold at 6.5 MeV is not yet sufficient
to exclude them. More precise conclusions can be reached in the future 
with the improvement of statistic and lowering of the threshold to 5 MeV.

%
%
We determined the values of $\Delta m^2$ and $\sin^2 2\theta$ 
needed in order to reproduce the present
event rates in HM, GALLEX/SAGE and SK by using models with
different values of $S_{17}$ (c.f.~Fig~\ref{plane}). 
For the best fit LMA, SMA and
VO-solutions (depending on $S_{17}$) we calculated the electron recoil
spectrum by convolving the
neutrino spectrum with the calculated survival probability, the
neutrino-electron scattering cross section and the energy resolution
function. 
%
%
Apparently, the spectrum in SK does not allow us to
discriminate among different values of $S_{17}$ (ii), but it can provide
important information to distinguish the different types of solutions (i).
We thus calculated the first and second electron
moments of the recoil electron energy distribution 
assuming a threshold of $5~\rm{MeV}$ and a energy scale
uncertainty $\delta=\pm100$~keV as in~\cite{bkr}.  Further 
information can in fact be extracted from the relative deviations
of the above two moments from the corresponding moments in the case 
of non-oscillating neutrinos $(\langle
E\rangle-\langle E\rangle_0)/\langle E\rangle_0$ and $(\langle
\sigma^2\rangle -\langle \sigma^2\rangle_0)/\langle
\sigma^2\rangle_0$ ~(the subscript ``0''
refers to the no-oscillation case.). 
As it is shown in Table~\ref{mome}, different solutions lead to different 
relative deviations of the first two spectral moments.

We note in particular that in the SMA case an increase of $S_{17}$ leads to an
increase of the relative deviation of both first and second
moments, while in the LMA case, one finds the opposite behavior
with a weaker relative variation.  A trend that is qualitatively
very similar to this one can also be observed for the sterile case.

\begin{table}[hb]
\caption{\label{mome} Fractional deviation from the no-oscillation
case of the first and second moment of the energy distribution of the
recoil electron in SK and SNO for active neutrinos.  The first two
columns refers to the SMA solution, the second ones to LMA, and the
last ones to VO.}
\smallskip
\begin{tabular}[2]{cccccccc}
\multicolumn{7}{c}{Super-Kamiokande}\\ $S_{17}$ & $\Delta E\;[\%]$ &
$\Delta \sigma^2\;[\%]$ & $\Delta E\;[\%]$ & $\Delta\sigma^2\;[\%]$
&$\Delta E\;[\%]$ & $\Delta\sigma^2\;[\%]$ \\
\noalign{\vskip3pt\hrule\vskip3pt} 14 & 0.98 & 3.38 & -0.37 & -1.51 &
5.90 & 6.88 \\ 19 & 1.41 & 4.98 & -0.49 & -1.58 & 3.32 & -1.64 \\ 23 &
1.56 & 5.61 & -0.12 & -0.32 & 0.77 & -9.80 \\ \hline\hline
\multicolumn{7}{c}{SNO}\\ $S_{17}$ & $\Delta E\;[\%]$ & $\Delta
\sigma^2\;[\%]$ & $\Delta E\;[\%]$ & $\Delta\sigma^2\;[\%]$&$\Delta
E\;[\%]$ & $\Delta\sigma^2\;[\%]$ \\
\noalign{\vskip3pt\hrule\vskip3pt} 14 & 1.31 & 2.04 & -0.15 & -0.36 &
3.06 & -19.7 \\ 19 & 2.17 & 2.91 & -0.55 & -0.72 & -0.21 & -21.3 \\ 23
& 2.62 & 3.53 & 0.02 & 0.31 & -3.10 & -24.2 \\
\end{tabular}
\end{table}

\subsection{Borexino}

\begin{figure}[hb]
\hbox to\hsize{\hss\epsfxsize=9cm\epsfbox{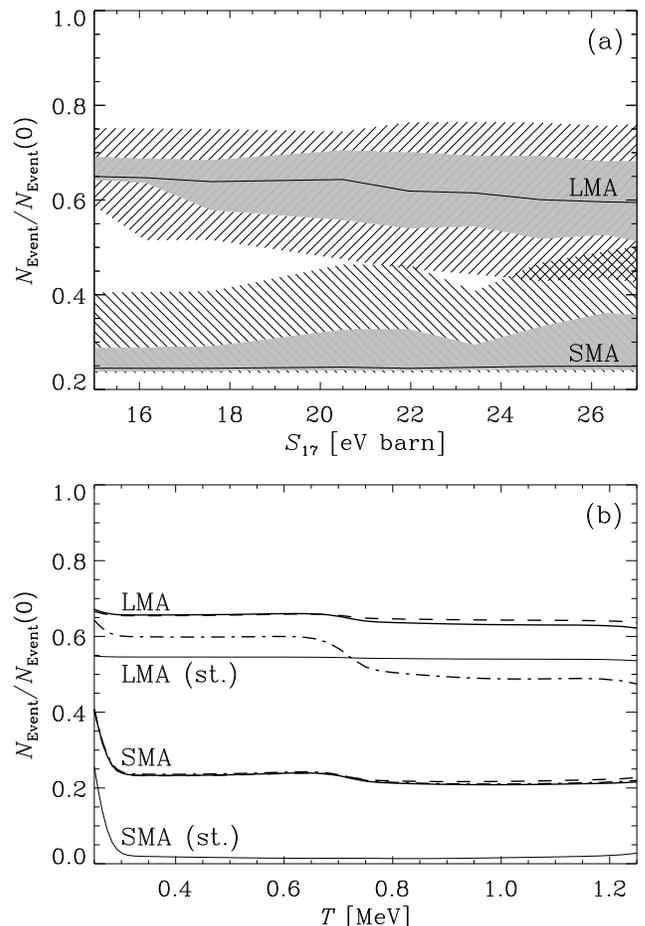}\hss}
\caption{(a) Event rates in Borexino normalized to the expected rates
from our standard solar model without oscillations for different
values of $S_{\rm 17}$. The shaded regions show the 1$\sigma$, the
hatched the 2$\sigma$ areas. (b) Recoil electron spectrum for the best
fit SMA and LMA solutions in sterile (st.) and non-sterile case. For
the latter $S_{\rm 17}= 15$ (dashed), 19 (solid) and 27 (dash-dotted)
in units of ${\rm eV\,barn}$.
\label{borexino}}
\end{figure}

The Borexino-experiment will measure mainly the ${\rm ^7Be}$-neutrinos
via neutrino-electron-scattering, therefore no significant information
can be obtained from this experiment about the value of $S_{17}$ (ii),
as the expected counting rate is independent of $S_{17}$
(Fig.~\ref{borexino}a). 

However, it is interesting to note that 
although the 1$\sigma$-regions of the SMA and LMA solution 
are well separated, at 2$\sigma$ level there is some
overlap. In this case it may also be possible that the 
measurement of the event rate will not be sufficient to discriminate these 
solutions unless the value of $S_{17}$ is quite low.

The expected recoil electron spectra are shown in Fig.~\ref{borexino}b
for the different types of solution.  The SMA solution shows a rise in
the signal at low energies, thus it is crucial to have good
statistical data just above the detector threshold of ${\rm 0.25~MeV}$. The
behaviour of the SMA solution is described by the typical shape of the
survival probability of electron neutrinos with varying energy
(``valley'' at intermediate energies).  This leads to an almost full
conversion of the $^7\rm{Be}$-neutrinos into $\nu_\tau, \nu_\mu$ or
$\nu_{\rm s}$, partial conversion of the ${\rm ^8B}$-neutrino and almost no
change of the $pp$-neutrinos.  In the case of the LMA-solution the
survival probability of $\nu_{\rm e}$ is almost constant for all the
energies. 
In the light of the present solar neutrino experiments results (total rates)
the MSW-SMA solution seems to be the most viable one for explaining the lack of 
$^7{\rm Be}$ and a reduction by a factor 2
of the $^8{\rm B}$ neutrinos. 

In the VO case ($10^{-11}$$\le$$\Delta m^2$$\le$$10^{-9}$) the
eccentric orbit of the earth leads to seasonal variations in the
neutrino flux due to the long oscillation length $l_{\rm V} \approx
2.48 E/\Delta m^2$ ($l_{\rm V}$ in m, $E$ in MeV, $\Delta m^2$ in
$eV^2$). Since 90\% of the $^7{\rm Be}$-neutrinos are emitted in a
monoenergetic line, this effect is more pronounced for these
neutrinos than for $pp$ and
$^8{\rm B}$-neutrinos, which are emitted in a continuous range of 
energies.  In the SMA and LMA solutions no seasonal variation appears, thus
Borexino should be able to discriminate between these cases and the
VO solution (i).

\subsection{SNO}
\begin{figure}[ht]
\hbox to\hsize{\hss\epsfxsize=9cm\epsfbox{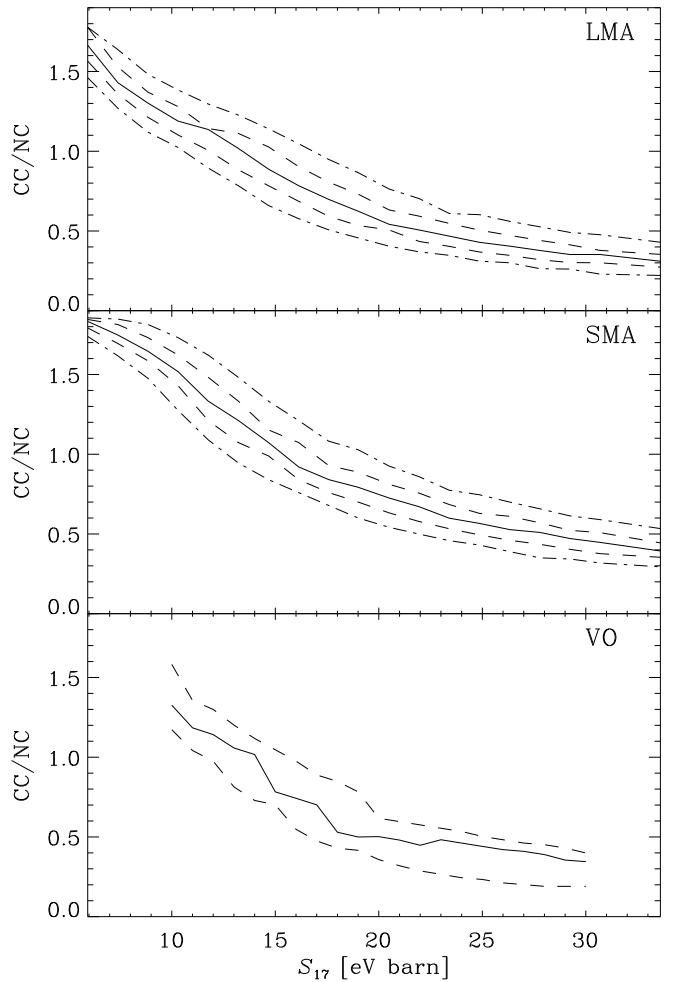}\hss}
\caption{(CC/NC)-ratio in SNO for solar models with varying values of
$S_{\rm 17}$. The dashed line shows the $1\sigma$, the dash-dotted
line the $2\sigma$-range.
\label{sno}}
\end{figure}

The SNO experiment will measures the recoil electron spectrum of the
reaction \[ \nu_e+d \rightarrow p+p+e^-\] and the ratio of the charged
to neutral current events (CC/NC). 
Using the neutrino fluxes of a solar model with a fixed value of
$S_{17}$, the expected  (CC/NC)-ratio is determined by
letting $\Delta m^2$ and $\sin^2 2\theta$ vary within the 68.4\%
and 95.4\% C.L.-region of the LMA, SMA or VAC-solution.
As shown in Fig.~\ref{plane}, varying $S_{17}$ change
the oscillation parameters which are able to reproduce the present
results of 
GALLEX/SAGE, SK and HM. These changes alter the expected
(CC/NC)-ratio, and thus provide an indirect dependence of the
(CC/NC)-ratio on $S_{17}$ (ii). The found relations are shown in
Fig.~\ref{sno}.
For the calulation of the (CC/NC)-ratio in SNO
we have used the energy resolution corresponding to a typical
statistics of 5000 CC events.
%
%

In the case of VO solution the $1\sigma$ level of uncertainty is
significantly larger than in the MSW solution case (Fig.~\ref{sno}). In the SMA
scenario it is possible to determine an effective constraint on
$S_{17}$ from the $1\sigma$-level strip of the (CC/NC)
ratio. For instance, from Fig.~\ref{sno} it can be inferred that 
a measurement of (CC/NC) of $\simeq 0.8$ would imply
\[S_{17}=19.0^{+2.0}_{-3.0}~{\rm keV\,barn},\]
if SMA turns out to be the solution of the solar neutrino puzzle.
In the VO case the limits are not very
stringent but they nevertheless provide independent constraints on the
allowed value of $S_{17}$.  However, this procedure is not very useful
for sterile neutrinos, because no sensible variation of the (CC/NC)
ratio occurs when $S_{17}$ is varied.

\begin{figure}[ht]
\hbox to\hsize{\hss\epsfxsize=9cm\epsfbox{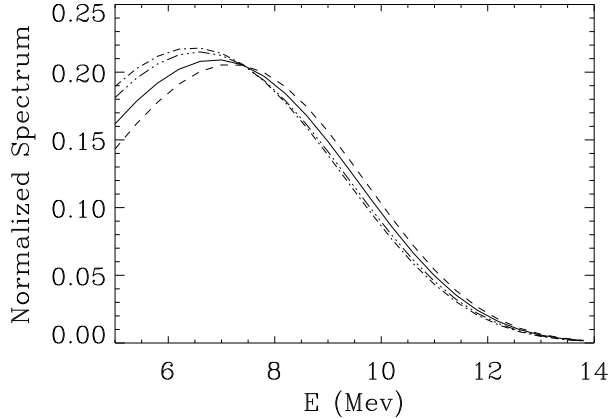}\hss}
\caption{Normalized electron energy spectra in SNO for active
neutrinos. The SMA solutions correspond to the solid and dashed lines
for $S_{17}=14~{\rm eV\,barn}$ and $S_{17}=23~{\rm eV\,barn}$
respectively.  The dash-double-dotted ($S_{17}=23~{\rm eV\,barn}$)
and dash-dotted lines ($S_{17}=14~{\rm eV\,barn}$) are for the LMA
solution.
\label{snospec}}
\end{figure}

The recoil electron spectrum provides additional information about the
type of the solution (i).  In particular we have employed a Gaussian
energy resolution function of width $\sigma_{10}=
1~{\rm MeV}$ at the electron energy $E_{\rm e}=10~{\rm MeV}$
as adopted in ~\cite{bkr}. 
For the best fit SMA and LMA solutions obtained from solar models with
different values of $S_{17}$, the expected electron energy spectrum in
SNO is shown in Fig.~\ref{snospec} for the case of active neutrinos.

The separation in the recoil electron spectra of both solutions is not
very pronounced, therefore these data alone may not be sufficient to
discriminate between LMA and SMA solution.  We remark that the overall
behaviour of the SMA and LMA solutions in SNO is very similar to the
one in SK, namely that the average energy of the recoil electrons
is higher in the SMA than in the LMA case for every value
of $S_{17}$ (see also Table~\ref{mome}).

In the sterile case the differences among various cases with altered
$S_{17}$ are much smaller (ii), and it is even more unlikely that any
significant variation in the spectra will be visible, neither in SNO
nor in SK.

\section{Conclusions}
We have investigated the influence of
$S_{11}$ on the sound speed and the small
spacing frequency differences by comparing the model predictions
with helioseismic data 
using up-to-date solar models. Moreover we
discussed the change  
in the allowed parameter space for SMA, LMA and VO solutions 
with varying $S_{17}$.  As shown in Section II the latest results from 
helioseismology suggest that the value of $S_{11}$ 
is slightly greater than the theoretically calculated one. However, since the
statistical significance is weak, we conclude that the limits
inferred from helioseismology and those derived from the theory
are consistent. The influence of the value of
$S_{11}$ on the solar neutrino flux is too small to alter the
resulting neutrino mixing parameters significantly. However, the proposed LENS
detector~\cite{rrr} can observe in principle a suppression of the
$pp$-neutrino flux and therefore it is reasonable to expect relevant
differences in the signal as function of the $S_{11}$ value.

The present experiments GALLEX/SAGE, HM and SK favour neutrino
oscillations as the solution to the solar neutrino deficit. Improved
statistics in SK and future experiments like Borexino and SNO will provide
powerful tools to support this solution.  We have calculated the
expected rates, electron moments, electron spectra or (CC/NC)-ratios
of the above experiments for the SMA, LMA and VO solution provided by
the present data. We expect that the combined data of the recoil
electron spectra in SK, SNO and Borexino enable us to discriminate
among these solutions.

Since the $^7{\rm Be}(p,\gamma)^8{\rm B}$ reaction has no influence on the solar
structure, it is impossible to get information about its strength from
helioseismology. Moreover the exact value for $S_{17}$ is crucial to
calculate the flux of the most energetic solar neutrinos, which are
measured in the SK and SNO experiments. The (CC/NC)-ratio
expected in SNO is sensitive to the $\phi_{\nu}(^8{\rm B})$ which is
directly related to the strength of $S_{17}$.

We conclude that the combination of SK, SNO and Borexino will be
useful to test the consistency of the value of $S_{17}$ found by
direct nuclear physics measurements with the combined analysis of 
theoretical models and neutrino experiments as described in Sections 
III and IV.  Of course, the whole analysis was done under the assumption of
neutrino-oscillations (either MSW or ``just so'') as solution to the
solar neutrino puzzle.  In the case of oscillations into sterile
neutrinos 
the strength of $^7{\rm Be}({\rm p},\gamma)^8{\rm B}$
does not leave any
signature in the future experiments. However, this
solution can be at least discriminated from the 
oscillations into active neutrinos by means of the behaviour of the 
(CC/NC) ratio.

\section*{Acknowledgments}

We are grateful to S.~Turck-Chi\`eze for useful discussions and for
allowing us to use a set of the GOLF data, and to J.~Christensen-Dalsgaard and
S.~Basu for providing us with the inverted sound speed profile 
derived from the GOLF+MDI data. We would also like to express our thanks to 
A.~Weiss, H.~M.~Antia, J.~N.~Bahcall and S.~Degli'Innocenti
for useful comments and advices.
The work of H.~S.~was partly supported by the ``Sonderforschungbereich
375-95 f\"ur Astrophysik'' der Deutschen For\-schungs\-ge\-mein\-schaft.
Furthermore A.~B.~acknowledges the INFN, Sezione di Catania and the
MPA for financial support, and thanks the scientists at
the MPA for their warm hospitality.

\end{document}